\documentclass[pra,twocolumn, showpacs]{revtex4}
\usepackage{graphicx}
\usepackage{amsmath}
\begin{document}

\bibliographystyle{unsrt}

\date{\today}

\title{Noiseless attenuation using an optical parametric amplifier}
\author{R.A. Brewster}
\author{I.C. Nodurft}
\author{T.B. Pittman}
\author{J.D. Franson}
\affiliation{Physics Department, University of Maryland Baltimore
County, Baltimore, MD 21250}

\begin{abstract}

The process of heralded noiseless amplification, and the inverse process of heralded noiseless attenuation, have potential applications in the context of quantum communications. Although several different physical implementations of heralded noiseless amplifiers have now been demonstrated, the research on heralded noiseless attenuators has been largely confined to a beam-splitter based approach. Here we show that an optical parametric amplifier (OPA), combined with appropriate heralding, can also serve as a heralded noiseless attenuator. The counterintuitive use of an optical amplifier as an attenuator is only possible due to the probabilistic nature of the device.

\end{abstract}


\maketitle

Although quantum optical signals cannot be deterministically amplified without adding noise \cite{caves1982}, it has recently been shown that  non-deterministic noiseless amplification is possible \cite{ralph2009}. Broadly speaking, the idea is related to the linear optics quantum computing paradigm \cite{knill2001}, where heralding signals are used to indicate the successful operation of a probabilistic device.  This type of heralded noiseless amplifier has led to a large number of implementations and applications (see, for example, \cite{xiang2010,ferreyrol2010,zavatta2011,gisin2010,curty2011,pitkanen2011,minar2012,scott2013,bruno2013,chrzanowski2014,ulanov2015,donaldson2015}).  Somewhat surprisingly, the inverse process of heralded noiseless attenuation is also significant \cite{micuda2012,gagatsos2014,zhao2017}.  In fact, Micuda {\em et.al.} have shown that noiseless amplification and noiseless attenuation can be combined to conditionally suppress the effects of loss in a quantum communication channel \cite{micuda2012}.

At the present time, heralded noiseless attenuators have only been studied using a beam-splitter based approach \cite{micuda2012,gagatsos2014}. This opens the question of whether other systems may be useful for this application.  In this brief paper, we show that a conventional optical parametric amplifier (OPA) --when combined with appropriate heralding-- can be used as a heralded noiseless attenuator. This unconventional use of an amplifier as an attenuator is somewhat related to the idea that the annihilation operator $\hat{a}$ can actually increase the average number of photons for certain states \cite{mizrahi2002}. It can also be viewed as an example of a more general equivalence between beam-splitters and OPA's in conditional measurements \cite{ban1997, barnett1999}.

\begin{figure}[b]
\includegraphics[width=3.25in]{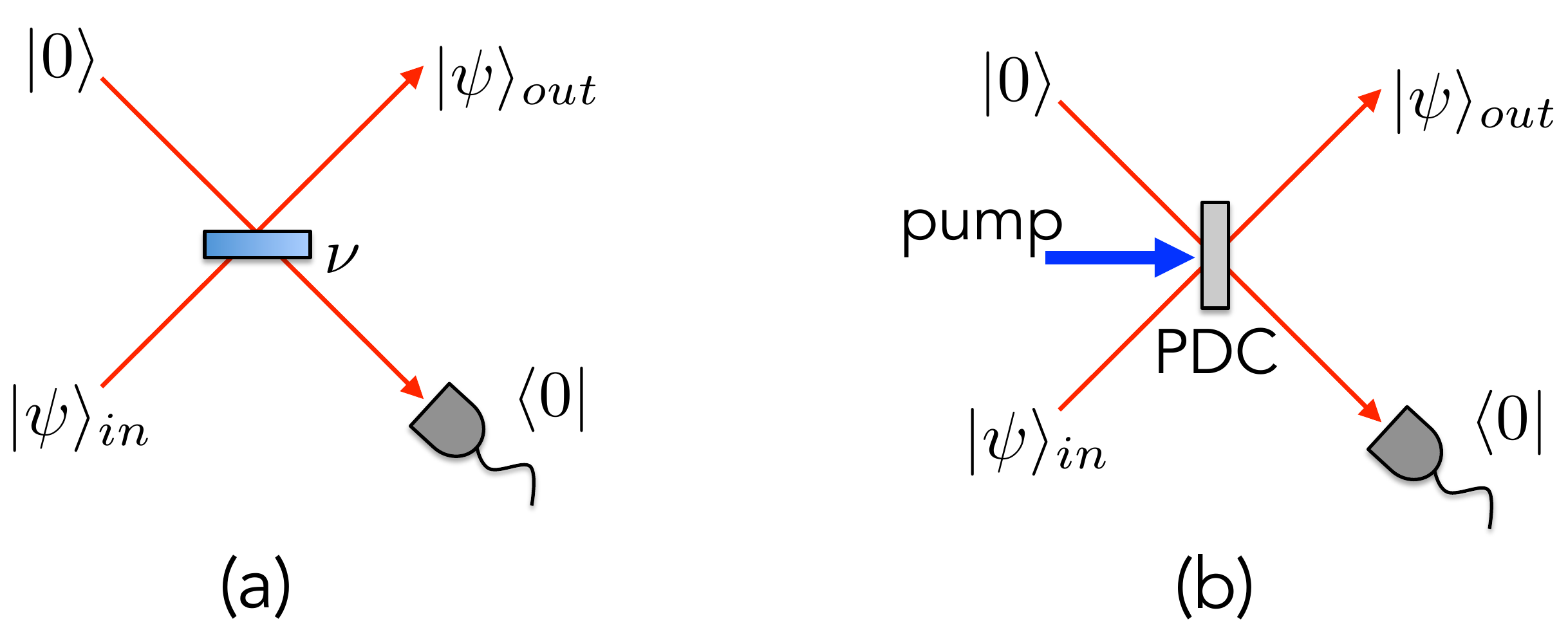}
\caption{(Color online) Two implementations of a heralded noiseless attenuator for coherent states. (a) shows a conventional beam-splitter based approach \protect\cite{micuda2012,gagatsos2014}. (b) shows the heralded noiseless parametric attenuator (NPA) formed by seeding a parametric down-converter (PDC) with the input state and vacuum. In both cases, the successful operation of the attenuator is heralded by detecting exactly zero photons in an auxiliary detector. The similar topologies of (a) and (b) highlights the similarities between multiphoton interference at a beam-splitter and stimulated emission \protect\cite{sun2007}.}
\label{fig:fig1}
\end{figure}

An overview of the basic idea is shown in Figure \ref{fig:fig1}. First, recall that the goal of the noiseless amplifer is to increase the amplitude of a coherent state by the transformation $|\alpha\rangle \rightarrow |g\alpha\rangle$, where $g\geq1$ is the gain and certain limits apply \cite{ralph2009,pandey2013}. In analogy, the goal of the noiseless attenuator is to reduce the amplitude as $|\alpha\rangle \rightarrow |\nu\alpha\rangle$, where $\nu\leq1$ is the attenuation parameter. In the beam-splitter based approach shown in Figure  \ref{fig:fig1}(a),  the input state and a vacuum state are mixed at a beam-splitter with amplitude transmittance $\nu < 1$, and the output state is only accepted when an auxiliary single-photon detector registers exactly zero photons \cite{micuda2012,gagatsos2014}. In this case, the heralded coherent state attenuation occurs because the post-selected probability amplitude of the Fock states $|n\rangle$ in the number-state expansion is reduced by a factor of $\nu^{n}$ \cite{micuda2012}.

In contrast, Figure \ref{fig:fig1}(b) shows the design of the noiseless attenuator considered in this paper. This basic arrangement of an OPA with heralding on one of the output modes has been extensively studied within the context of general conditional quantum state engineering by multiphoton addition \cite{sperling2014}. Here we specifically focus on the case of heralding on zero photons to realize a noiseless attenuator, which we refer to as a heralded noiseless parametric attenuator (NPA). 

 As shown in Figure \ref{fig:fig1}(b), the input state and a vacuum state are used to seed the signal and idler inputs of a parametric down-converter (PDC), and the output state is only accepted when the auxiliary  detector registers exactly zero photons in the idler mode. In this case, the desired attenuation essentially occurs because higher number Fock states $|n\rangle$  are increasingly more likely to stimulate the emission of a signal-idler photon pair in the PDC process \cite{sun2007}. Consequently as $|n\rangle$ increases, the more likely it is to produce an idler photon, which renders it less likely to ``survive'' the heralding process. This results in a corresponding decrease in the relative amplitudes of these terms in the coherent state expansion, and a corresponding overall attenuation of the coherent state amplitude.

In order to explicitly calculate this attenuation, we use a relatively simple method based on the time evolution (two-mode squeezing) operator for an OPA given by
\begin{equation}
\hat{S}(r)={{e}^{r(\hat{a}\hat{b}-{{{\hat{a}}}^{\dagger }}{{{\hat{b}}}^{\dagger }})}},
\label{eq:squeezeOp}
\end{equation}
where $r=\kappa t$, $\kappa$ is the coupling between the pump and the signal and idler modes, $t$ is the time, and $\hat{a}$ and $\hat{b}$ are the annihilation operators for the signal and idler modes respectively \cite{agarwalBook}. It has been shown by Schumaker and Caves that equation (\ref{eq:squeezeOp}) can be rewritten in a factored form in the following way \cite{schumaker1985, caves2012}
\begin{equation}
\hat{S}(r)=\frac{1}{g}{{e}^{-\sqrt{{{g}^{2}}-1}{{{\hat{a}}}^{\dagger }}{{{\hat{b}}}^{\dagger }}/g}}{{g}^{-({{{\hat{a}}}^{\dagger }}\hat{a}+{{{\hat{b}}}^{\dagger }}\hat{b})}}{{e}^{\sqrt{{{g}^{2}}-1}\hat{a}\hat{b}/g}},
\label{eq:factoredOp}
\end{equation}
where $g=\cosh(r)$ is the gain of the OPA.

The idler mode will be assumed to initially be in its vacuum state ${{\left| 0 \right\rangle }_{i}}$. The unitary transformation $\hat{S}(r)$ followed by the heralding process gives the projection ${{\left| 0 \right\rangle }_{i}}\left\langle  {{0}_{i}} \right|\hat{S}(r){{\left| \alpha  \right\rangle }_{s}}{{\left| 0 \right\rangle }_{i}}$.  In evaluating this expression, the last exponential on the right-hand side of equation (\ref{eq:factoredOp}) reduces to
\begin{equation}
{{e}^{\sqrt{{{g}^{2}}-1}\hat{a}\hat{b}/g}}{{\left| \alpha  \right\rangle }_{s}}{{\left| 0 \right\rangle }_{i}}={{\left| \alpha  \right\rangle }_{s}}{{\left| 0 \right\rangle }_{i}},
\label{eq:identityExample}
\end{equation}
since $\hat{b}|0\rangle_i=0$. In the same way, the first exponential on the right-hand side of equation (\ref{eq:factoredOp}) reduces to the identity operator when acting to the left, since $\langle0|_i\hat{b}^\dagger=0$.

Combining equations (\ref{eq:factoredOp}) and (\ref{eq:identityExample}) with the usual expression for a coherent state gives the heralded state of the output mode as
\begin{equation}
\begin{split}
  & {{\left\langle  0 \right|}_{i}}\hat{S}(r){{\left| \alpha  \right\rangle }_{s}}{{\left| 0 \right\rangle }_{i}} \\ 
 & =\frac{1}{g}{{e}^{-{{\left| \alpha  \right|}^{2}}/2}}\sum\limits_{n=0}^{\infty }{\frac{{{\alpha }^{n}}}{\sqrt{n!}}{{g}^{-{{{\hat{a}}}^{\dagger }}\hat{a}}}}{{\left| n \right\rangle }_{s}} \\ 
 & =\frac{1}{g}{{e}^{-{{\left| \alpha  \right|}^{2}}/2}}\sum\limits_{n=0}^{\infty }{\frac{1}{\sqrt{n!}}{{\left( \frac{\alpha }{g} \right)}^{n}}}{{\left| n \right\rangle }_{s}}. \\ 
\end{split}
\label{eq:calculation}
\end{equation}
The unnormalized state $|\psi\rangle_\text{out}$ of the output mode can be rewritten as
\begin{equation}
{{\left| \psi  \right\rangle }_{\text{out}}}=\frac{1}{g}{{e}^{-({{g}^{2}}-1){{\left| \alpha  \right|}^{2}}/2{{g}^{2}}}}{{\left| \alpha /g \right\rangle }_{s}}.
\label{eq:cohOut}
\end{equation}
If we define the attenuation parameter by  $\nu =1/g$, then it can be seen from equation (\ref{eq:cohOut}) that the NPA results in the transformation $|\alpha \rangle \to |\nu \alpha \rangle $.   The probability ${{P}_{s}}$ of success is just the square of the coefficient in front of the state in Eq. (5), which gives 
\begin{equation}
{{P}_{s}}=\frac{1}{{{g}^{2}}}{{e}^{-({{g}^{2}}-1){{\left| \alpha  \right|}^{2}}/{{g}^{2}}}}.
\label{eq:cohProb}
\end{equation}
Equation (\ref{eq:cohOut}) is somewhat remarkable in that the heralding process converts a coherent state into an attenuated coherent state rather than a more complicated result \cite{footnote1}.

Simply passing a coherent state through a beam-splitter (or absorbing filter) without heralding would also give an attenuated coherent state. That is not the case, however, for more complicated superposition states such as a Schr\"{o}dinger cat state $|\psi\rangle=(1/\sqrt{2})(|\alpha\rangle+|-\alpha\rangle)$, for example. Although each component would have a reduced amplitude, the coherence between the two states would be eliminated by which-path information left in the environment \cite{kirby2013, franson2017}. The heralded noiseless attenuator described here or in references  \cite{micuda2012,gagatsos2014} avoids decoherence of that kind by post-selecting on a single state of the ``environment", namely the idler mode in this case.

This property also allows the NPA to coherently attenuate the most basic superposition state: a single-rail qubit \cite{footnote2}. It can be shown that applying the operator $\hat{S}(r)$ to a Fock state $|n\rangle $ after proper heralding on a zero-photon detection event in the idler mode gives
\begin{equation}
{{\left\langle  0 \right|}_{i}}\hat{S}(r){{\left| n \right\rangle }_{s}}{{\left| 0 \right\rangle }_{i}}=\frac{1}{{{g}^{n+1}}}{{\left| n \right\rangle }_{s}}.
\label{eq:numOut}
\end{equation}
Applying equation (\ref{eq:numOut}) to, for example, the single rail qubit $|\psi\rangle_\text{in}=(1/\sqrt{2})(|0\rangle+|1\rangle)$ would give an output state of the form
\begin{equation}
{{\left| \psi  \right\rangle }_{\text{out}}}=\frac{1}{g\sqrt{2}}\left( {{\left| 0 \right\rangle }_{s}}+\frac{1}{g}{{\left| 1 \right\rangle }_{s}} \right).
\label{eq:qubitOut}
\end{equation}
It can be seen from equation (\ref{eq:qubitOut}) that the single rail qubit is also attenuated by a factor $\nu=1/g$, while the probability of success is
\begin{equation}
{{P}_{s}}=\frac{{{g}^{2}}+1}{2{{g}^{4}}}.
\label{eq:numProb}
\end{equation}

In summary, we have introduced the heralded noiseless parametric attenuator (NPA) as an alternative to the beam-splitter based approach to noiseless attenuation \cite{micuda2012,gagatsos2014}. A heralding signal is essentially used to convert the gain $g$ of a conventional OPA into an attenuation factor $\nu=1/g$. Larger gain values result in higher attenuation, but a rapidly decreasing probability of success (cf. equations (\ref{eq:cohProb}) and (\ref{eq:numProb})).  From a practical point of view, the experimental realization of an NPA appears to be feasible with existing technology. The effective gain values for state-of-the-art single-pass single-mode waveguide-based PDC sources can be remarkably large \cite{harder2013}, and the use of currently available single-photon detectors with low dark count rates and high detection efficiencies would help overcome the difficulties associated with heralding on  ``zero photons" \cite{eisaman2011}.

As is the case with a heralded noiseless amplifier \cite{ralph2009}, we note that the NPA studied here does not add or subtract photons, but simply reweights the coefficients in the Fock state expansion of the input state. It is exactly this reweighting that allows the combination of a noiseless attenuator and a noiseless amplifier to suppress loss in quantum communication channels \cite{micuda2012}. In addition, we emphasize that this unconventional use of an amplifier as an attenuator is only possible due to the probabilistic (non-unitary) nature of the heralding process. 

In this brief paper, we have specifically considered the action of the NPA on coherent states and single-rail qubit inputs. The transformation of more exotic states is also interesting, and may have further applications in robust quantum communications \cite{brewster2017}. 

{\bf Acknowldegments:} This work was supported in part by the National Science Foundation under grant No. 1402708. We acknowledge fruitful conversations with G.T. Hickman and H. Lamsal.


\end{document}